\documentclass[aps,twocolumn,superscriptaddress]{revtex4}

\usepackage[version=3]{mhchem} % Formula subscripts using \ce{}
\usepackage{graphicx}% Include figure files
\usepackage{epstopdf}
\usepackage{dcolumn}% Align table columns on decimal point
\usepackage{bm}% bold math
\begin{document}

\title{Two-Dimensional Magnetotransport in a Black Phosphorus Naked Quantum Well}

\author{V. Tayari}
\affiliation{Department of Electrical and Computer Engineering, McGill University, Montr\'eal, Qu\'ebec, H3A 2A7, Canada}
\author{N. Hemsworth}
\affiliation{Department of Electrical and Computer Engineering, McGill University, Montr\'eal, Qu\'ebec, H3A 2A7, Canada}
\author{I. Fakih}
\affiliation{Department of Electrical and Computer Engineering, McGill University, Montr\'eal, Qu\'ebec, H3A 2A7, Canada}
\author{A. Favron}
\affiliation{Department of Physics, Universit\'e de Montr\'eal, Montr\'eal, Qu\'ebec, H3C 3J7, Canada}
\author{E. Gaufr\`es}
\affiliation{Department of Chemistry, Universit\'e de Montr\'eal, Montr\'eal, Qu\'ebec, H3C 3J7, Canada}
\author{G. Gervais}
\affiliation{Department of Physics, McGill University, Montr\'eal, Qu\'ebec, H3A 2T8, Canada}
\author{R. Martel}
\affiliation{Department of Chemistry, Universit\'e de Montr\'eal, Montr\'eal, Qu\'ebec, H3C 3J7, Canada}
\author{T. Szkopek}
\email{thomas.szkopek@mcgill.ca}
\affiliation{Department of Electrical and Computer Engineering, McGill University, Montr\'eal, Qu\'ebec, H3A 2A7, Canada}

\date{\today}

\begin{abstract}
Black phosphorus (bP) is the second known elemental allotrope with a layered crystal structure that can be mechanically exfoliated down to atomic layer thickness. We have fabricated bP naked quantum wells in a back-gated field effect transistor geometry with bP thicknesses ranging from $6\pm1$ nm to $47\pm1$ nm. Using an encapsulating polymer superstrate, we have suppressed bP oxidation and have observed field effect mobilities up to 600 cm$^2$/Vs and on/off current ratios exceeding $10^5$. Importantly, Shubnikov-de Haas (SdH) oscillations observed in magnetotransport measurements up to 35 T reveal the presence of a 2-D hole gas with Schr\"odinger fermion character in an accumulation layer at the bP/oxide interface. Our work demonstrates that 2-D electronic structure and 2-D atomic structure are independent. 2-D carrier confinement can be achieved in layered semiconducting materials without necessarily approaching atomic layer thickness, advantageous for materials that become increasingly reactive in the few-layer limit such as bP.
\end{abstract}

\maketitle

Layered two-dimensional (2-D) materials have undergone a renaissance since the development of mechanical exfoliation techniques \cite{geimPNAS}. Black phosphorus (bP) is a layered material (Fig. 1a) with van der Waals interlayer bonding \cite{morita}, and is the only elemental allotrope other than graphene that is presently known to be a 2-D material. Recent work has shown that bP can be exfoliated down to the atomic limit \cite{zhang, xia, tomanek, martel, ozyilmaz, gomez}. In bulk form, bP is a narrow gap semiconductor with a 0.3 eV direct bandgap \cite{keyes}, which grows to a $\sim$2 eV bandgap in the atomic monolayer limit \cite{morita}, ideal for application to transistors \cite{schwierz}. Ambipolar conduction, mobilities approaching $\sim$1000 cm$^2$/Vs, and anisotropic conductivity have been demonstrated \cite{zhang, xia, tomanek}, leading to a revitalized interest in bP \cite{ye}. 

Interestingly, it was long ago observed \cite{morita} that despite the weak van der Waals bonding between the 2-D atomic layers of bP, the effective mass for electron (hole) motion between planes is remarkably light at 0.13$m_0$ (0.28$m_0$) \cite{narita}. Exfoliated bP layers are thus effectively naked quantum wells with a low charge trap density at the bP surface due to the absence of broken covalent bonds and the simultaneous delocalization of charge carriers across atomic layers due to the light effective mass. The high electronic quality of the naked bP surface, requiring no passivation, is rare among semiconductors.

\begin{figure}
    \includegraphics [width=2.75in]{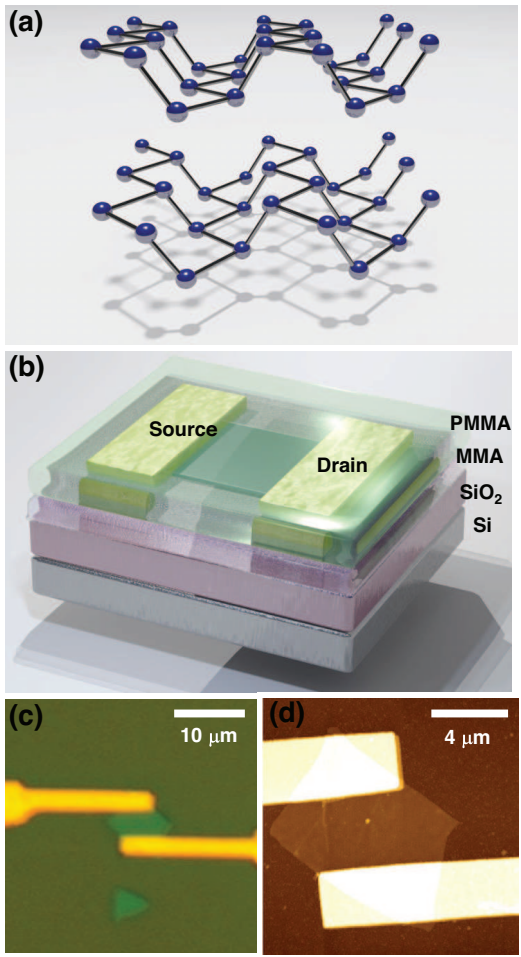}
    \caption{\label{} \textbf{Structure of black phosphorus FETs}. \textbf{a}, The bP crystal structure is composed of puckered honeycomb layers with an interlayer distance of 5.24 \AA. \textbf{b}, Three-dimensional schematic view of a bP FET with oxidized silicon back-gate and an encapsulating layer of MMA and PMMA. \textbf{c}, Optical image of the encapsulated device A. \textbf{d}, AFM image of device A with encapsulating layer removed. The bP thickness is $11.5\pm1$ nm ($22\pm2$ atomic layers).     }
\end{figure}

In our work, we have fabricated field effect transistors (FETs) with exfoliated bP layers ranging in thickness from $6\pm1$ nm to $47\pm1$ nm ($11\pm2$ to $90\pm2$ atomic layers). Despite being the most stable allotrope of phosphorus, bP suffers from photo-oxidation in a reaction that proceeds faster as atomic film thickness is approached \cite{martel}. The deleterious effects of photo-oxidation were mitigated by using bP layers thicker than a few atomic layers, by encapsulating the bP in a polymer superstrate, and by minimizing exposure to oxygen, water and visible light. We have measured the electronic transport properties of bP FETs over the temperature range of 0.3 K to 300 K, including the measurement of Shubnikov-de Haas (SdH) oscillations at magnetic fields up to 35 T. The observed SdH oscillations indicate the presence of a 2-D hole gas in an accumulation layer as in conventional semiconductor heterostructures, demonstrating that 2-D carrier confinement can be achieved in bP of $\sim 90$ atomic layer thickness which are much less susceptible to photo-oxidation than few-layer bP. 

\section{Results}

Ultra-thin bP samples were prepared by mechanical exfoliation from bulk bP crystals using a polydimethylsiloxane (PDMS) stamp technique previously reported \cite{martel}. The sample substrates were degenerately doped Si wafers, with 300 nm of dry thermal oxide to allow rapid optical identification of bP flakes and back-gating over a wide temperature range. To protect bP FETs against degradation, 300 nm of copolymer (methyl methacrylate) and 200 nm of polymer (polymethyl methacrylate) were deposited. A schematic of the bP FET structure is shown in Fig.1b. The polymer layer forms a water impermeable superstrate suppressing oxidation. An optical reflection image under white light illumination of a typical encapsulated bP FET (device A) is shown in Fig. 1c.  Upon completion of electron transport measurements, described in detail below, the encapsulating polymer was removed with acetone and the layer thickness was measured by atomic force microscopy (AFM) within a glove box. An AFM image of an unencapsulated bP FET (device A) is shown in Fig. 1d. The bP layer thickness of this representative device was determined to be 11.5$\pm1$ nm ($22\pm2$ atomic layers). Most importantly, the bP surface is free of the surface roughening that arises from oxidation \cite{ozyilmaz, martel, gomez, hersam}, despite exposure to ambient conditions. Encapsulation with a PMMA/MMA superstrate, similar to encapsulation with parylene \cite{martel} or AlO$_x$ \cite{hersam}, was thus found to be an effective means to suppress photo-oxidation of multi-layer bP.

\begin{figure}
    \includegraphics [width=3.15in]{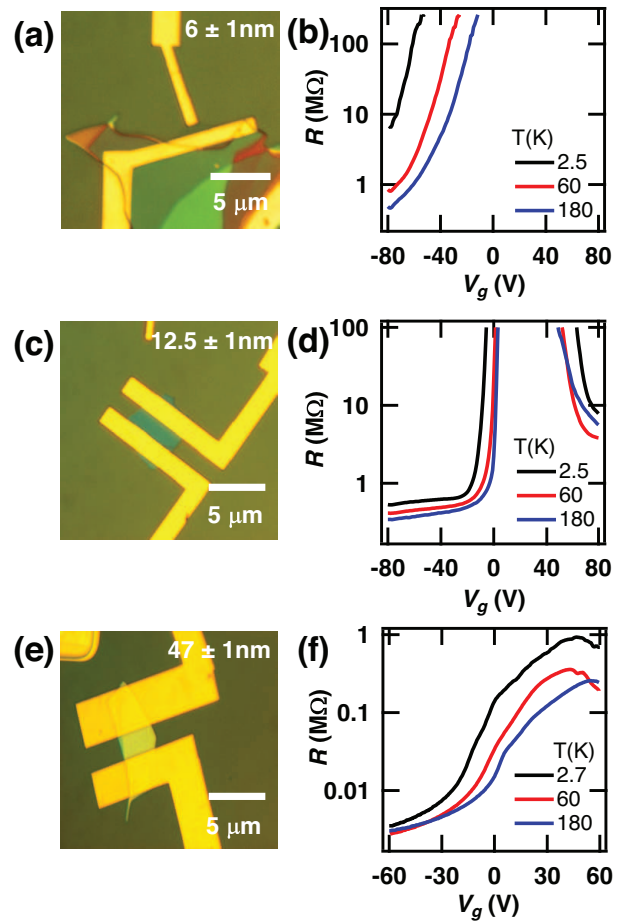}
    \caption{\label{} \textbf{Black phosphorus FET characterization }. \textbf{a},\textbf{c},\textbf{e}, Optical images of device B, C and D with increasing thickness of $6\pm1$, $12.5\pm1$ and $47\pm1$ nm, respectively. \textbf{b},\textbf{d},\textbf{f}, Source-drain resistance $R$ as a function of gate voltage $V_g$ at different temperatures for the devices shown in \textbf{a}, \textbf{c} and \textbf{e} respectively. The thickest device D exhibits a resistance two orders of magnitude smaller than devices B and C. }
\end{figure}

Charge transport was investigated in over 40 bP FETs, of which three representative samples are shown Fig. 2. The thinnest bP FET measured (device B) was 6$\pm$1 nm ($11\pm 2$ atomic layers) thick, displayed in Fig. 2a. The source-drain resistance $R$ measured by ac lock-in technique at $\sim10$ Hz versus gate voltage $V_g$ with a source drain bias of $V_{sd}=10$ mV is illustrated in Fig. 2b at different temperatures. The resistance exceeded our measurement limit for electrons, while strong insulating behaviour, $\partial R / \partial T < 0$, was observed for holes. A sample of 12.5$\pm$1 nm ($24\pm2$ atomic layers) thickness (device C) is displayed in Fig. 2c, with measured resistance in Fig. 2d. Ambipolar conduction is observed, with a threshold for hole conduction at a gate voltage $V_g \sim -15$V and a threshold for electron conduction at $V_g \sim 65$V at low temperature. The bP FET channel is thus slightly hole doped. A weaker temperature dependence of the resistance $R$ at high hole density is observed as compared to the thinner device B. The thickest bP FET measured (device D) was 47$\pm$1 nm ($90\pm2$ atomic layers) thick, displayed in Fig. 2e. The measured resistance $R$, plotted in Fig. 2f, is two orders of magnitude smaller than in the aforementioned devices. The thickest device displays a low on/off current ratio of $10^2$ at 180K. In contrast, on/off current ratios exceeding $10^5$ can be achieved in thin samples. The field effect hole mobility, $ \mu_{FE}=(L/W)\cdot \partial(1/R)/\partial(C_g V_g)$ where $C_g$ is the gate capacitance per unit area for a channel of length $L$ and width $W$, reaches $\approx600$ cm$^2/$Vs at $V_g\approx-80$ V for device D at room temperature, and is amongst the highest field effect mobility observed in our experiments. The electrodes in our samples are not aligned with the crystallographic axes of the bP layers, and thus our observed mobility is an average over the anisotropic transport properties of bP \cite{xia}.

To further elucidate the nature of hole conduction in bP, the magnetoresistance of our highest mobility device D was measured at low temperature, $T=0.3$ K in a magnetic field up to $B$ = 35 T. The two-point source-drain  resistance $R$ of device D was measured by ac lock-in technique with a 10 nA bias current, and is plotted in Fig. 3a versus magnetic field oriented normal to the bP atomic planes at different gate voltages $V_g$. The gate voltages were selected to induce a hole gas of varying density in the bP FET. The magnetoresistance exhibits a weak localization peak at low field $B<1$T, a smooth positive magnetoresistance background, and Shubnikov-de Haas (SdH) oscillations at fields exceeding $\sim15$ T. The SdH oscillations were analyzed by fitting the resistance $R(B)$ to a parabolic function, as shown in Fig. 3a, to subtract the smooth magnetoresistance background. The resultant oscillating magnetoresistance component $\Delta R$ is plotted versus $1/B$ in Fig. 3b. The low carrier density and high magnetic field regime of our experiments leads to the approximation $\Delta R \propto \Delta R_{xx} \propto \Delta \sigma_{xx}$, from which a Landau-Kosevich (LK) form of SdH oscillations follows, $\Delta R = R_D(B) \cos[ 2 \pi ( B_F/B + 1/2 + \beta ) ]$ where $B_F$ is the magnetic frequency, $\beta$ is the normalized Berry phase and $R_D(B)$ is the damping factor arising from hole scattering \cite{shoenberg}. An example of a best fit of the SdH oscillations to the LK form is indicated with a dashed line in Fig. 3b.

\begin{figure}
    \includegraphics [width=2.75in]{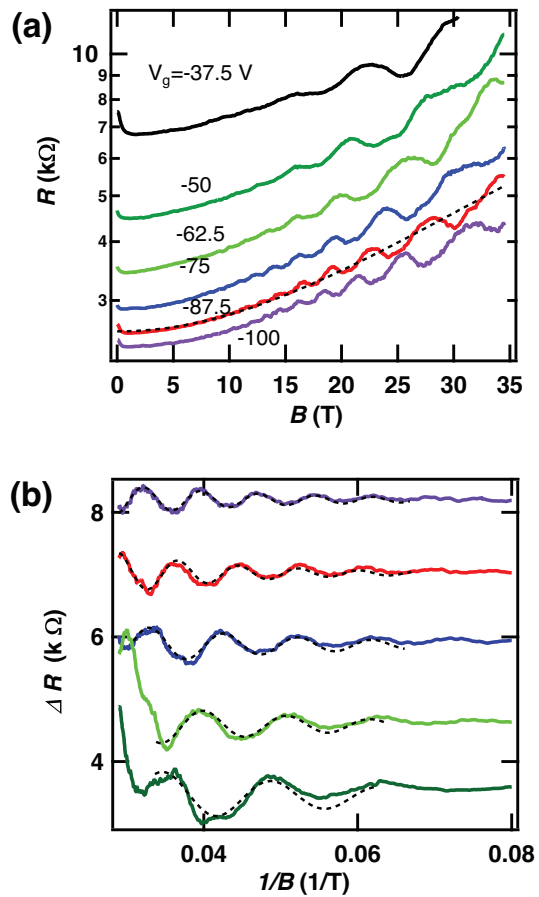}
    \caption{\label{} \textbf{Shubnikov-de Haas oscillations}. \textbf{a}, The measured resistance $R$ shown in Fig. 2e on a log scale as a function of applied magnetic field $B$ at different gate voltages at T=$0.3$ K. A weak localization peak at low field, a slowly varying positive magnetoresistance, and SdH oscillations are observed. The slowly varying positive magnetoresistance has a parabolic form, an example of a best-fit being shown by the dashed line. \textbf{b}, Subtracting the parabolic magnetoresistance background, the oscillating resistance $\Delta R$ is plotted as a function of $1/B$, with vertical offsets for clarity. The SdH oscillations were fit to the Landau-Kosevitch formula, indicated by dashed lines. The quality of the fit improves with increasing hole density. }
\end{figure}

We first consider the scattering damping factor $R_D(B) = R_0 \exp( -2 \pi^2 m^* k_B T_D / \hbar e B )$, where $T_D$ is the Dingle temperature and $m^* = \sqrt{m_x m_y} = 0.22 m_0$ is the effective mass appropriate for in-plane cyclotron motion with a value determined from cyclotron resonance experiments \cite{narita}. At the highest hole density, corresponding to gate voltage $V_g = -100$ V, we find $T_D = 16\pm1$ K from which we estimate the hole scattering time $\tau = \hbar / 2 \pi k_B T_D  = 75\pm5$ fs and a magnetoresistive mobility $\mu_{MR} = e \tau / m^* = 640\pm40$ cm$^2$/Vs. The magnetoresistive mobility is in good agreement with the independently measured field effect mobility.

The Landau levels (LLs) at the origin of SdH oscillations can be further analyzed with a Landau fan diagram of LL index $N$ versus $1/B$, illustrated in Fig. 4a. The LL index $N$ corresponds to the $N^{th}$ minimum in $\Delta R$ versus $1/B$. The half integer index $N+1/2$ corresponds to the $N^{th}$ maximum in $\Delta R$. The LK fit to the SdH oscillations of Fig. 3b at each gate voltage were used to determine the minima and maxima presented in the fan diagram of Fig. 4a. Notably, our experiments closely approach the quantum limit, with an LL index as small as $N + 1/2 = 2.5$ being observed. At each gate voltage, the magnetic frequency $B_F$ determined by the LK fit corresponds to the slope of the Landau fan diagram $B_F=\delta N/\delta (1/B)$. The Berry phase $\beta$ corresponds to the LL index intercept at $1/B = 0$ in the Landau fan diagram. The Berry phase determined at each gate voltage is summarized in Fig. 4b, and is consistent with a trivial phase $\beta = 0$ of Schr\"odinger fermions. This is theoretically expected from the direct gap at the time reversal invariant momentum of the $Z$-point of the Brillouin zone, which forces the Berry curvature to vanish in the absence of spin splitting induced by a large spin-orbit coupling, as expected here for bP. 

\begin{figure}
    \includegraphics [width=2.75in]{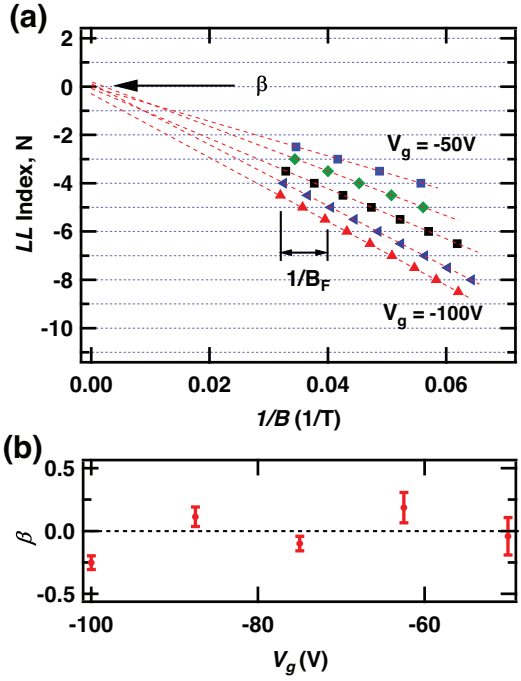}
    \caption{\label{} \textbf{Landau fan diagram analysis. } \textbf{a},  The Landau fan diagram of LL index $N$ versus $1/B$ at different gate voltages extracted from Landau-Kosevich analysis of the SdH oscillations in Fig 3b. The SdH frequency $B_F$ and Berry phase $\beta$ are extracted from the slope and intercept of the diagram. \textbf{b}, The Berry phase $\beta$ versus gate voltage $V_g$ is consistent with $\beta = 0$ for Schr\"odinger fermions. }
\end{figure}

Further information about the hole gas in the bP FET can be determined by comparison of the free charge density per unit area versus the charge density induced by field effect $n_{\mathrm{FE}} = C V_g / e$ where $C = 11.5$ nF/cm$^2$ is the gate capacitance per unit area of 300 nm of SiO$_2$. In the case of a 2D hole gas occupying a single valley at Z with unbroken spin degeneracy, the free charge density is $n_{\mathrm{free}} = 2 B_F\cdot e/h$. Spin degeneracy appears unbroken throughout the SdH oscillations of Fig 3b, likely due to the strong disorder broadening of the LL's indicated by the large Dingle temperature $T_D = 16\pm1$ K. The free charge versus gated charge is plotted in Fig. 5a, along with a linear best fit that corresponds to a 78\% gate efficiency. The linearity and proximity to ideal gate behaviour is consistent with the presence of a 2D hole gas within the 47 nm thick bP FET. In contrast, the carrier density versus magnetic frequency for a 3-D hole gas is given by $n_{\mathrm{3D}}=n_{\mathrm{2D}}/t =8\sqrt{\pi}/3\cdot(B_F\cdot e/h)^{3/2}$ where $t$ is the effective thickness of the hole gas\cite{shoenberg, kohler}. The measured free carrier density versus charge density induced by field effect does not agree with a 3-D model unless a hole gas thickness of $t\approx3$ nm is used, in which case 2-D quantum confinement effects must be taken into account. In other words, the observed magnetic frequency variation versus gate voltage indicates the presence of a 2-D hole gas rather than a 3-D hole gas.

\begin{figure}
    \includegraphics [width=2.75in]{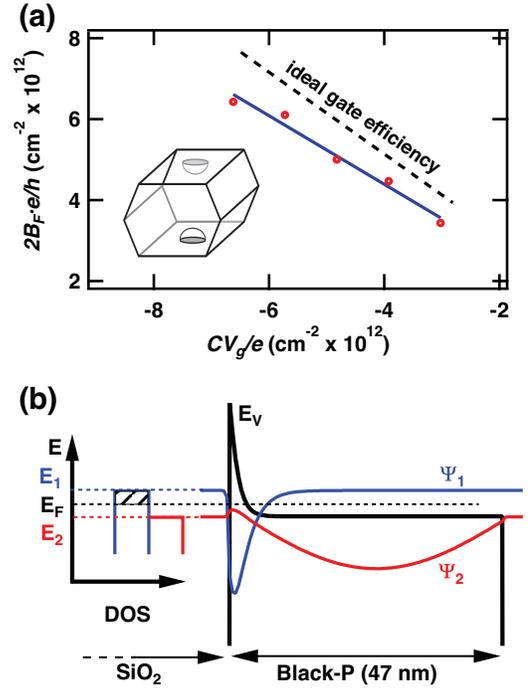}
    \caption{\label{} \textbf{Free carrier density analysis} \textbf{a}, The free carrier density $2 B_F \cdot e/h$ (for a single, spin degenerate hole valley at Z) versus the charge density induced by field effect $C V_g / e$ (red circles). A linear fit (blue) corresponds to a 78\% gate efficiency. \textbf{b} Self-consistent Schr\"odinger-Poisson calculation of the ground state wavefunction $\Psi_1$ ($E_1-E_F = 13$ meV) and first excited state wavefunciton $\Psi_2$ ($E_F-E_2 = 15$ meV) for holes, and the corresponding valence band edge $E_V$, at a gate bias of $V_g = -50$ V. A single 2D sub-band is occupied. }
\end{figure}

We estimated the 2-D quantum confinement of holes at the bP/oxide interface with a self-consistent Schr\"odinger-Poisson calculation using an effective mass band theory. Estimates of the occupied and un-occupied 2D sub-band wavefunctions, valence band edge and density of states are illustrated in Fig. 5c at a gate bias of $V_g = -50$ V. The hole gas in the wide bP quantum well is similar to the 2D accumulation (or inversion) layer induced in Si FETs at an Si / SiO$_2$ interface. The strong electric field, of order $\sim 0.1$ V/nm, applied to induce a hole gas within the bP results in a 2D sub-band with a wavefunction $\Psi_1$ that is tightly confined to an rms width of $2.7$ nm ($\sim5-6$ atomic layers). At $V_g=-50$ V, Schr\"odinger-Poisson calculation places the Fermi level $E_F$ 13 meV below the ground state sub-band edge $E_1$, but the first excited sub-band $E_2$ lies 15 meV below the Fermi level and is thus energetically inaccessible to holes at $T$ = 0.3 K. The bP/oxide interface is expected to be rich in charge traps, and is consistent with the modest mobilities and deviation from ideal gate efficiency observed for the 2D hole gas confined to the bP/oxide interface in our work. However, the saturation of in-plane chemical bonding within bP likely leads to a low density of dangling bonds that would otherwise lead to Fermi level pinning and suppression of the electric field effect.

We have thus shown that two distinct aspects of 2-D physics can co-exist in one material system. On the one hand, the 2-D nature of covalent bonding in bP is favourable for mechanical exfoliation of thin layers and results in a low dangling bond density at naked bP surfaces. On the other hand, the field effect can be used to induce a charge accumulation layer in a single 2-D electronic sub-band as in conventional semiconductor heterostructures and quantum wells. 2-D electronic behaviour can be accessed in multi-layer devices, and this degree of freedom is anticipated to be useful in the future development of 2-D electronics, including van der Waals heterostructures that combine multiple layered materials together. In the case of layered semiconductor materials whose stability decreases as the few-layer limit is approached such as bP, multi-layer stacks can be used while retaining 2-D electronic behaviour.

\section{Methods}

\subsection{Device Fabrication}

The source material for bP preparation were 99.998\% purity bP crystals from Smart Elements (Vienna, Austria). Mechanical exfoliation of bP layers was performed using a PDMS stamp technique within a nitrogen glove box environment to minimize exposure to water and oxygen, as previously reported \cite{martel}. The sample substrates were degenerately doped Si wafers, with 300 nm of chlorinated dry thermal oxide. The substrates were pre-patterned with metal alignment marks, and were annealed at 150$^\circ$ C for 15 min to desorb water prior to bP exfoliation. Optical reflection microscopy with red light (using a 580 nm long pass filter) was performed to identify thin bP flakes while minimizing the effects of photo-oxidation. Conventional electron beam lithography and metal deposition were used to define Ti/Au (5 nm/ 80 nm) contacts on bP flakes, with care taken to minimize simultaneous exposure to water, oxygen and visible light. Once fabricated, the bP FETs were encapsulated in a glove box environment by spin-coating 300 nm of copolymer (methyl methacrylate) and 200 nm of polymer (polymethyl methacrylate) followed by an annealing step at 170$^\circ$ C for 15 minutes.

\subsection{Quasi-DC Characterization}

Initial charge transport experiments on the bP FETS were performed under quasi-dc bias in a vacuum probe station with a semiconductor parameter analyzer. The characteristics of the representative bP FET device A are displayed in Fig. 6. The source-drain current $I$ was measured at fixed source-drain bias $V_{sd}$ versus gate voltage $V_g$ swept in both directions. Most devices exhibited ambipolar conduction, and an on/off current ratio that increased as temperature was decreased from 300 K to 77 K. As seen in Fig. 6, hysteresis was also observed to rapidly decrease as temperature decreased. Gate leakage current was simultaneously monitored in all experiments, and never exceeded 10\% of the minimum source-drain current. The source-drain current versus bias voltage $V_{sd}$ was also measured for all devices as shown in Fig. 6b. 

\begin{figure}
    \includegraphics [width=2.75in]{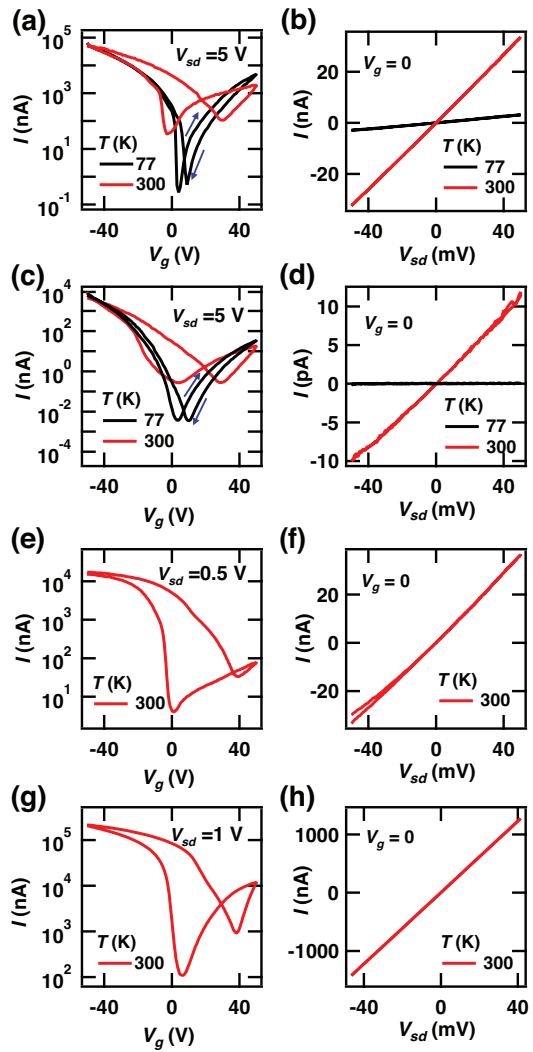}
    \caption{\label{} \textbf{Quasi-dc FET characterization} \textbf{a},\textbf{c},\textbf{e},\textbf{g}, The source-drain current $I$ of devices A, B, C, D respectively at fixed source-drain bias $V_{sd}$ versus gate voltage $V_g$. Room temperature and 77 K measurements are presented for devices A and B, showing a decrease in hysteresis and increase in on/off ratio at 77 K . \textbf{b},\textbf{d},\textbf{f},\textbf{h}, The source-drain current $I$ of devices A, B, C, D respectively with a gate bias of $V_g=0$ V versus source-drain bias $V_{sd}$. Ohmic behaviour is observed.  }
\end{figure}

\subsection{AC Characterization}

Following quasi-dc characterization, devices were selected for characterization over a wider temperature range. Samples were mounted on fiber glass chip-carriers and electrical contact made by wire bonding. A variable temperature insert in a helium cryostat was used to measure source-drain resistance down to temperatures of 2.5 K using standard ac lock-in measurement with a 10 mV voltage bias. Magnetotransport measurements were conducted in a helium-3 cryostat in a resistive magnet cell at the National High Magnetic Field Laboratory (Tallahassee, Florida). During all magnetotransport measurements, the sample was immersed in a helium-3 bath at 300 mK, and source-drain resistance was again measured using standard ac lock-in techniques.

\subsection{Raman Spectroscopy}

\begin{figure}
    \includegraphics [width=2.75in]{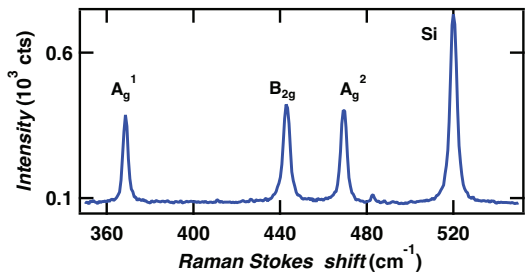}
    \caption{\label{} \textbf{Raman spectroscopy} Raman Stokes shift of encapsulated bP FET device A measured with a 532 nm laser pump. A silicon substrate peak is observed as well as the $A_g^1$, $B_{2g}$ and $A_g^2$ Raman modes of bP. }
\end{figure}

Raman spectroscopy was performed once all electron transport experiments were completed, to minimize the effect of photo-oxidation on the electronic quality of the bp FETs. Raman spectroscopy was performed with the sample in a vacuum cell using a custom built instrument with a laser pump at 532 nm and a numerical aperture of NA = 0.55. The pump fluence is estimated to be 20 kW/cm$^2$. The resolution of the Raman spectrometer is $\pm0.2$ cm$^{-1}$. A representative Raman Stokes spectrum of device A is shown Fig. 7. The strong peak at 520 cm$^{-1}$ originates from the Si substrate \cite{parker67}. The three peaks observed at 368.7, 442.8 and 469.2 cm$^{-1}$ correspond to the $A_g^1$, $B_{2g}$ and $A_g^2$ Raman modes of bP as reported in studies of bulk, single crystal bP \cite{kaneta}.

\subsection{Atomic Force Microscopy}

Upon completion of Raman spectroscopy, the PMMA/MMA superstrate layers were removed from bP FETs by immersion in acetone in a glove box environment. AFM was performed within the same glove box with a ThermoMicroscopes Auto Probe CP in order to determine the thickness of the bP layers. AFM images were acquired in intermittent imaging mode with 85\% damping with Al-coated Si cantilever probes (tip radius $< 10$ nm, spring constant 25-75 N/m). AFM images of devices B, C, D are shown in Fig 8, from which thicknesses of $6\pm1$, $12.5\pm1$ and $47\pm1$ nm were determined, respectively.

\begin{figure}
    \includegraphics [width=3.25in]{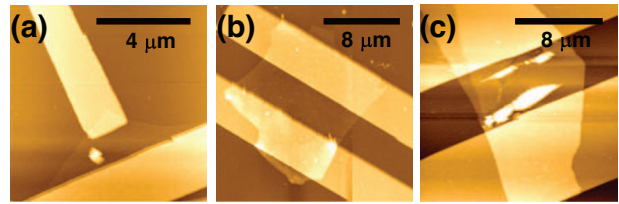}
    \caption{\label{} \textbf{AFM images} \textbf{a},\textbf{b},\textbf{c}, AFM images of unencapsulated bP FET devices B, C, D with respective thicknesses of  $6\pm1$, $12.5\pm1$ and $47\pm1$ nm.  }
\end{figure}

\subsection{Schr\"odinger-Poisson Analysis}

The nature of the 2D hole gas accumulating at the bP/oxide was modelled with self-consistent Schr\"odinger-Poisson calculations. The calculations were performed in one dimension by iterative solution of an effective mass Schr\"odinger equation and Poisson's equation for the mean-field electrostatic potential of the hole density. Calculations were performed with a bP effective hole mass of $0.28m_0$, a bP dielectric constant $6.1\epsilon_0$ \cite{berger}, and 3 eV potential barriers at the bP surface as a model for a hard potential barrier into the oxide substrate or polymer superstrate. Fermi-Dirac statistics at 3K were used for hole population, and the field effect was modelled by including an applied electric field. \\

\section{Acknowledgements}

We thank N. Doiron-Leyraud and M.O. Goerbig for useful discussion, N. Tang for assistance with AFM, D. Cardinal for assistance with wire-bonding, L. Engel for assistance with high-field measurements, and G. Jones, J. Jaroszynski and T.P. Murphy for outstanding technical support for high-field measurements. This work was funded by NSERC, CIFAR, FRQNT, RQMP and the CRC program. A portion of this work was performed at the National High Magnetic Field Laboratory which is supported by NSF Cooperative Agreement No. DMR-0084173, the State of Florida, and the DOE. 

Upon completion of this work, the authors have learned of similar results obtained by Likai Li et al., arXiv:1411.6572 [cond-mat.mes-hall].

\end{document}